\documentclass[letterpaper]{jpconf}
\usepackage{graphicx}
\begin{document}
\title{Light quark mass effects in the chromomagnetic moment}

\author{S. Bekavac$^{(a)}$, A. Grozin$^{(a,b)}$, D. Seidel$^{(c)}$ and M. Steinhauser$^{(a)}$}

\address{(a) Institut f\"ur Theoretische Teilchenphysik, Universit\"at Karlsruhe (TH),\\
\hspace{1.3em} Karlsruhe Institute of Technology (KIT), 76128 Karlsruhe, Germany\\
(b) Budker Institute of Nuclear Physics, Novosibirsk 630090, Russia\\
(c) Department of Physics, University of Alberta, Edmonton, Alberta, Canada T6G 2G7
}

\begin{abstract}
We present the three-loop QCD corrections to the quark chromomagnetic moment including two
different nonzero masses. This is a necessary ingredient to obtain the corresponding
corrections to the chromomagnetic coefficient in the Heavy Quark Effective Theory (HQET)
Lagrangian.
\end{abstract}

\section{Introduction}

The anomalous magnetic moment of the electron and the muon are among the most precisely
measured observables in particle physics. Comparing the theoretical and experimental
predictions for the muon magnetic moment there is currently a $3.4\sigma$
discrepancy~\cite{Passera:2008hj} with the Standard Model (SM) which makes this observable very
interesting at the moment. We calculate finite light quark mass contributions to the
chromomagnetic moment of quarks, and obtain, as a byproduct, the corresponding corrections to
the above mentioned observables and can confirm the results of
Refs.~\cite{Barbieri:1974nc,Laporta:1993ju,Laporta:1992pa}. Another byproduct of our
calculation is the anomalous magnetic moment of heavy quarks, the bottom quark in particular,
where we include the effect of a finite charm quark mass. The magnetic moment of quarks has not
yet been measured experimentally, however, for the bottom and the lighter quarks there are
upper limits from LEP1 data~\cite{Escribano:1993xr}. Due to the lack of space in these
proceedings we will present analytic results for this observable in Ref.~\cite{BGSS}.

Whereas the anomalous magnetic moments of fermions are physical observables, the chromomagnetic
moment is not. Nevertheless, it plays a crucial role in HQET, where it enters the matching
coefficient of the chromomagnetic interaction operator~\cite{Grozin:2007fh}. The one-loop
correction to the chromomagnetic moment has been obtained in
Refs.~\cite{Eichten:1989zv,Falk:1990pz}. In Refs.~\cite{Amoros:1997rx,Czarnecki:1997dz}, the
two-loop calculation has been performed, whereas light quark mass effects to this order have
been obtained in Ref.~\cite{Davydychev:1998si}. An estimation of higher order corrections has
then been given in Ref.~\cite{Grozin:1997ih} and the three-loop correction with one mass scale
was finalized in Ref.~\cite{Grozin:2007fh}. In the latter the aforementioned matching
coefficient of HQET is almost trivially obtained from the chromomagnetic moment. In the case
with two mass scales, additional diagrams have to be calculated in the effective theory to
match it to full QCD. We will present the matching coefficient in Ref.~\cite{BGSS} and restrict
the following discussion to the chromomagnetic moment. The results given in these
proceedings have been published in Ref.~\cite{PHDStefan}.

\section{Calculation of the chromomagnetic moment}

\begin{figure}
\begin{center}
\includegraphics[width=0.9\textwidth]{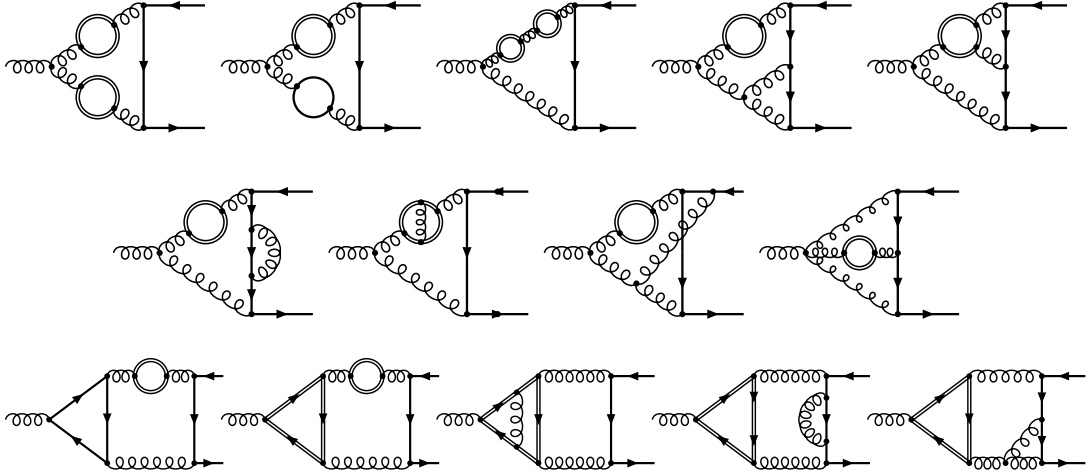}
\end{center}
\caption{\label{fig::diag}
Sample diagrams contributing to the quark chromomagnetic moment. Double solid lines denote
light quarks, whereas solid and curly lines denote heavy quarks and gluons, respectively. Note
that we use the background field method for the external gluon.}
\end{figure}

To calculate the chromomagnetic moment we have to consider the quark--anti-quark--gluon vertex
in the background-field formalism in QCD. We consider the effect of a nonzero light quark mass
at the three-loop level to this quantity. Sample diagrams which have to be calculated are
depicted in Fig.~\ref{fig::diag}. When both the quark and anti-quark are on the (renormalised)
mass shell and have physical polarisations, the vertex $\Gamma^\mu_a = \Gamma^\mu t_a$ can be
decomposed into two form factors,
\begin{equation}
  \Gamma^\mu = \gamma^\mu\, F_1(q^2)
    - \frac{i}{2 m_Q} \sigma^{\mu\nu} q_\nu F_2(q^2) \,,
  \label{eq::tensor}
\end{equation}
where $q = p_1 - p_2$ is the gluon momentum and $p_1$ and $p_2$ are the momenta of the quark
and anti-quark, respectively.

The anomalous chromomagnetic moment is given by $\mu_c = Z_2^{\rm OS} F_2(0)$, where $Z_2^{\rm
OS}$ is the quark wave function renormalisation constant in the on-shell scheme. The total
quark colour charge is given by $Z_2^{\rm OS} F_1(0) = 1$. Thus, $F_1(0)$ is the inverse of the
on-shell wave function renormalisation constant, which has been calculated to three-loops
including light quark masses in Ref.~\cite{Bekavac:2007tk}. Therefore, the calculation of
$F_1(0)$ provides a strong check on the correctness of our result.

All Feynman diagrams are generated with {\tt QGRAF}~\cite{Nogueira:1991ex} and the various
topologies are identified with the help of {\tt q2e} and {\tt
exp}~\cite{Harlander:1997zb,Seidensticker:1999bb}. In a next step the reduction of the various
functions to so-called master integrals has to be achieved. For this step we use the so-called
Laporta method~\cite{Laporta:1996mq,Laporta:2001dd} which reduces the three-loop integrals to
27 master integrals. We use the implementation of Laporta's algorithm in the program {\tt
Crusher}~\cite{PMDS}. It is written in {\tt C++} and uses {\tt GiNaC}~\cite{Bauer:2000cp} for
simple manipulations like taking derivatives of polynomial quantities. In the practical
implementation of the Laporta algorithm one of the most time-consuming operations is the
simplification of the coefficients appearing in front of the individual integrals. This task is
performed with the help of {\tt Fermat}~\cite{fermat} where a special interface has been used
(see Ref.~\cite{Tentyukov:2006ys}). The main features of the implementation are the automated
generation of the integration-by-parts (IBP) identities~\cite{Chetyrkin:1981qh}, a complete
symmetrisation of the diagrams and the possibility to make use of a multiprocessor environment.
As we need the form factors at zero momentum transfer all occurring master integrals are
on-shell propagator-type integrals. They have been calculated using different analytical and
numerical methods, see Refs.~\cite{Bekavac:2007tk,Bekavac:2009gz} for details.
To calculate the colour factors, we have used the program described in
Ref.~\cite{vanRitbergen:1998pn}.

\section{Results}

\begin{figure}
\begin{center}
\includegraphics[width=0.49\textwidth]{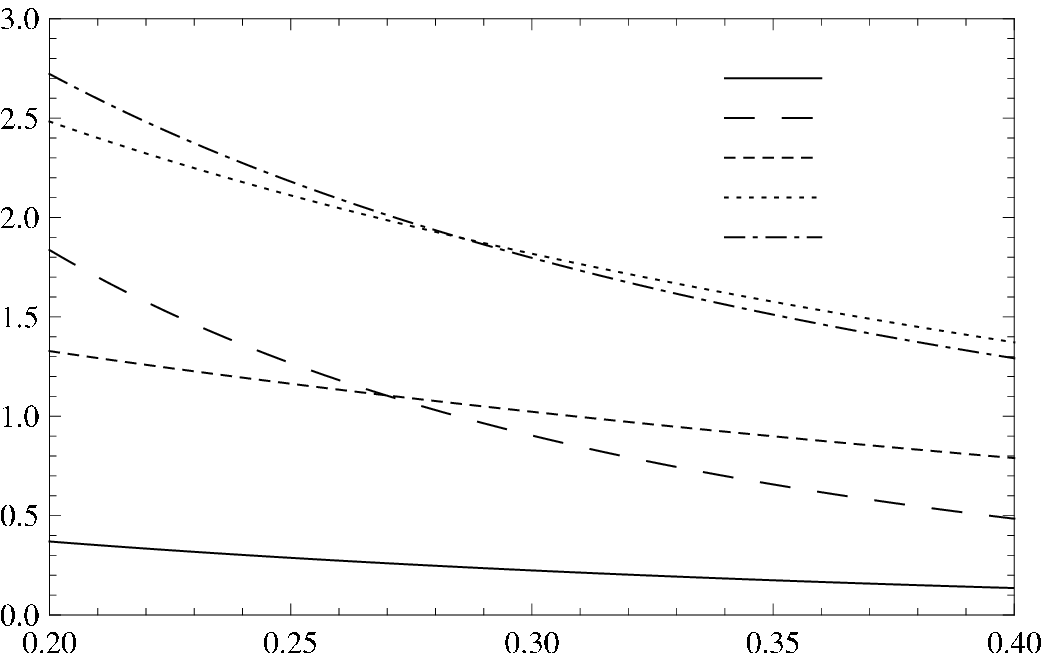}
\includegraphics[width=0.49\textwidth]{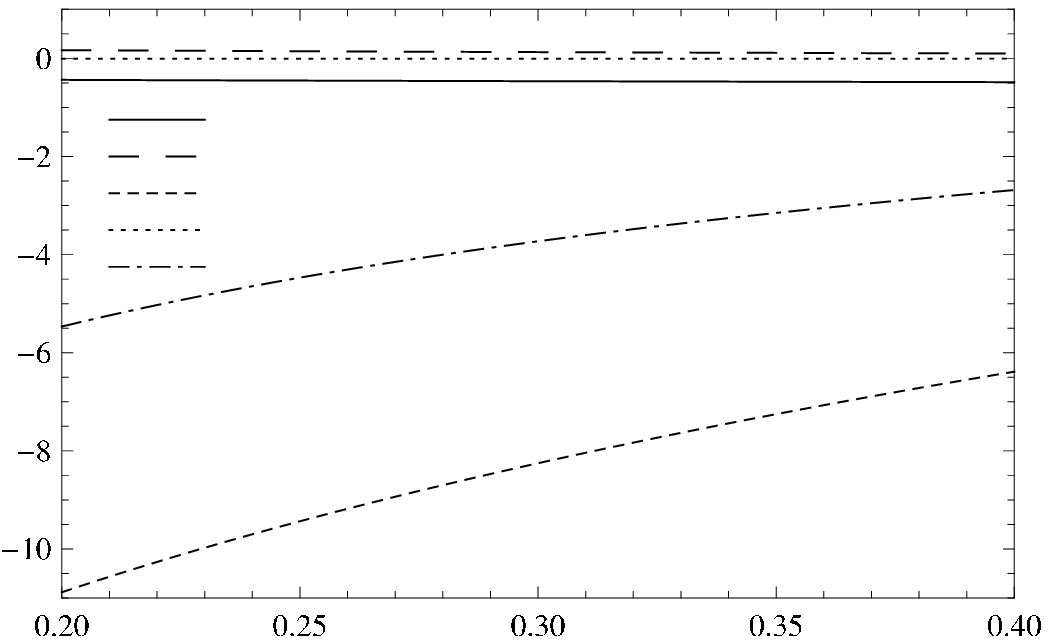}
\put(-270,121){\tiny $C^{FMM}$}
\put(-270,112.5){\tiny $C^{AMM}$}
\put(-270,104){\tiny $C^{FML}$}
\put(-270,95.5){\tiny $C^{AML}$}
\put(-270,87){\tiny $C^{DFM}$}
\put(-176,109){\tiny $C^{FFM}$}
\put(-176,101){\tiny $C^{AMH}$}
\put(-176,93){\tiny $C^{FAM}$}
\put(-176,85){\tiny $C^{FMH}$}
\put(-176,77){\tiny $C^{AAM}$}
\end{center}
\caption{\label{fig::plot}
Contributions of ${\cal O}(\varepsilon^0)$ from the different colour structures in
(\ref{eq::C}) as a function of the mass ratio $x$. Note the different scales of the two
diagrams.}
\end{figure}

We write the chromomagnetic moment in the form
\begin{eqnarray}
   \mu_c(\mu)  =
   1 +
   \sum_{n=1}^\infty\left(\frac{\alpha_s(\mu)}{\pi}\right)^n
   \left(\frac{e^{\gamma_E}}{4\pi}\right)^{-n\varepsilon} C^{(n)},
\end{eqnarray}
where $\gamma_E=0,57721...$ is the Euler-Mascheroni constant. $\alpha_s$ denotes the strong
coupling constant with $n_f=n_l+n_m+n_h$ active flavours ($n_l$, $n_m$ and $n_h$ are the
number of massless, light and heavy quarks, respectively). In practise, $n_m$ and $n_h$ will be
equal to one, but we keep them explicitly in our results in order to track the various
classes of diagrams in the result. We further decompose the three-loop contribution containing
light quarks with nonzero mass into its colour structures
\begin{eqnarray}
   C^{(3)}_{n_m} &=& 
   C^{FFM} C_F^2 T_F n_m +
   C^{AAM} C_A^2 T_F n_m +
   C^{FAM} C_A C_F T_F n_m +
   C^{FMM} C_F T_F^2 n_m^2 \nonumber\\
&& +C^{FLM} C_F T_F^2 n_l n_m +
   C^{FMH} C_F T_F^2 n_h n_m +
   C^{AMM} C_A T_F^2 n_m^2 +
   C^{ALM} C_A T_F^2 n_l n_m \nonumber\\
&& +C^{AMH} C_A T_F^2 n_h n_m +
   C^{DFM} \frac{d_F^{abcd}d_F^{abcd} n_m}{C_F N_F},
\label{eq::C}
\end{eqnarray}
where $C_F= (N_c^2 - 1)/(2N_c)$ and $C_A= N_c$ are the eigenvalues of the quadratic Casimir
operators of the fundamental and adjoint representation for the SU$(N_c)$ colour group,
respectively. In the case of QCD we have $N_c=3$ and $T_F = 1/2$. The dimension of the
fundamental representation is given by $N_F=N_c$. The symmetrised trace
of four generators in the fundamental representation is given by
$d_F^{abcd} d_F^{abcd} = (N_c^2-1)(N_c^4-6N_c^2+18)/(96N_c^2)$.
We present our results at the renormalisation scale $\mu=M_h$, where $M_h$ is the pole mass of
the heavy quark. The nonzero pole parts of the various contributions in (\ref{eq::C}) are given
by
\begin{eqnarray}
   C^{AMM} &=& \frac{\ln^2 x}{9\varepsilon} + {\cal O}(\varepsilon^0),\\
   C^{AML} &=& \frac{\ln x}{18\varepsilon^2} 
   - \frac{1}{\varepsilon}\left(\frac{\ln^2 x}{18} +\frac{13 \ln x}{108}+\frac{\pi
^2}{432}\right)
   + {\cal O}(\varepsilon^0),\\
   C^{AAM} &=& -\frac{\ln x}{9\varepsilon^2}
   + \frac{1}{\varepsilon}\left(
   \frac{5 \ln^2 x}{72}+\frac{155 \ln x}{432}
   -\frac{\pi ^2}{432}-\frac{137}{864}+\frac{\pi ^2}{16}\,x
   \right.\\
 && \left. + x^2 \left(-\frac{\ln^2 x}{16} +\frac{\ln x}{4}-\frac{\pi
   ^2}{96}-\frac{3}{16}\right)-\frac{\pi^2}{96}\,x^3\right)
   + {\cal O}(\varepsilon^0),\nonumber\\
   C^{FAM} &=& \frac{1}{\varepsilon}\left(
   \frac{5\ln x}{24}-\frac{235}{576}+\frac{\pi ^2}{16}\,x
   +x^2 \left(\ln x+\frac{3}{4}\right)-\frac{5\pi^2}{16}\, x^3
   \right) + {\cal O}(\varepsilon^0).
\end{eqnarray}
The contribution to $C^{AAM}$ and $C^{FAM}$ is presented as a series expansion up to third
order in the quark mass ratio $x=M_m/M_h$, where $M_m$ is the pole mass of the light quark. The
results for the finite parts of the different colour structures are given in graphical form
in Fig.~\ref{fig::plot} for $0.2<x<0.4$, which is relevant for charm mass effects in the
chromomagnetic moment of the bottom quark. The mass dependence of the bottom quark in the
chromomagnetic moment of the top quark can safely be neglected and the results at $x=0$
from~\cite{Grozin:2007fh} can be used. The analytic results including the renormalisation
scale dependence will be given in~\cite{BGSS}.

\ack
This work was supported by the DFG through SFB/TR 9 and the Graduiertenkolleg
"Hochenergiephysik und Teilchenastrophysik", the Science and Engineering Research Canada and
the Alberta Ingenuity Fund. We would like to thank Jan Piclum for fruitful discussions about
the project and Vladimir A. Smirnov for cooperation in the calculation of the diagrams.


\vspace{2em}

\begin{thebibliography}{99}
\bibitem{Passera:2008hj}
  M.~Passera, W.~J.~Marciano and A.~Sirlin,
  AIP Conf.\ Proc.\  {\bf 1078} (2009) 378
  [arXiv:0809.4062 [hep-ph]].

\bibitem{Barbieri:1974nc}
  R.~Barbieri and E.~Remiddi,
  Nucl.\ Phys.\  B {\bf 90} (1975) 233.

\bibitem{Laporta:1993ju}
  S.~Laporta,
  Nuovo Cim.\  A {\bf 106} (1993) 675.

\bibitem{Laporta:1992pa}
  S.~Laporta and E.~Remiddi,
  Phys.\ Lett.\  B {\bf 301} (1993) 440.

\bibitem{Escribano:1993xr}
  R.~Escribano and E.~Masso,
  Nucl.\ Phys.\  B {\bf 429} (1994) 19
  [arXiv:hep-ph/9403304].

\bibitem{BGSS} S. Bekavac, A. Grozin, D. Seidel and M. Steinhauser, in preparation

\bibitem{Grozin:2007fh}
  A.~G.~Grozin, P.~Marquard, J.~H.~Piclum and M.~Steinhauser,
  Nucl.\ Phys.\  B {\bf 789} (2008) 277
  [arXiv:0707.1388 [hep-ph]].

\bibitem{Eichten:1989zv}
  E.~Eichten and B.~R.~Hill,
  Phys.\ Lett.\  B {\bf 234} (1990) 511.

\bibitem{Falk:1990pz}
  A.~F.~Falk, B.~Grinstein and M.~E.~Luke,
  Nucl.\ Phys.\  B {\bf 357} (1991) 185.

\bibitem{Amoros:1997rx}
  G.~Amoros, M.~Beneke and M.~Neubert,
  Phys.\ Lett.\  B {\bf 401} (1997) 81
  [arXiv:hep-ph/9701375].

\bibitem{Czarnecki:1997dz}
  A.~Czarnecki and A.~G.~Grozin,
  Phys.\ Lett.\  B {\bf 405} (1997) 142
  [Erratum-ibid.\  B {\bf 650} (2007) 447]
  [arXiv:hep-ph/9701415].

\bibitem{Davydychev:1998si}
  A.~I.~Davydychev and A.~G.~Grozin,
  Phys.\ Rev.\  D {\bf 59} (1999) 054023
  [arXiv:hep-ph/9809589].

\bibitem{Grozin:1997ih}
  A.~G.~Grozin and M.~Neubert,
  Nucl.\ Phys.\  B {\bf 508} (1997) 311
  [arXiv:hep-ph/9707318].

\bibitem{PHDStefan}
  S. Bekavac,
  PhD-Thesis, Karlsruhe, January 2009; Shaker Verlag, Aachen, May 2009.

\bibitem{Bekavac:2007tk}
  S.~Bekavac, A.~Grozin, D.~Seidel and M.~Steinhauser,
  JHEP {\bf 0710} (2007) 006
  [arXiv:0708.1729 [hep-ph]].

\bibitem{Nogueira:1991ex}
  P.~Nogueira,
  J.\ Comput.\ Phys.\  {\bf 105} (1993) 279.

\bibitem{Harlander:1997zb}
  R.~Harlander, T.~Seidensticker and M.~Steinhauser,
  Phys.\ Lett.\  B {\bf 426} (1998) 125
  [arXiv:hep-ph/9712228].

\bibitem{Seidensticker:1999bb}
  T.~Seidensticker,
  arXiv:hep-ph/9905298.

\bibitem{Laporta:1996mq}
  S.~Laporta and E.~Remiddi,
  Phys.\ Lett.\  B {\bf 379} (1996) 283
  [arXiv:hep-ph/9602417].

\bibitem{Laporta:2001dd}
  S.~Laporta,
  Int.\ J.\ Mod.\ Phys.\  A {\bf 15} (2000) 5087
  [arXiv:hep-ph/0102033].

\bibitem{PMDS}
  P.~Marquard and D.~Seidel,
  unpublished.

\bibitem{Bauer:2000cp}
  C.~Bauer, A.~Frink and R.~Kreckel,
  arXiv:cs.sc/0004015.

\bibitem{fermat} R.~H.~Lewis, Fermat's User Guide,
  http://www.bway.net/\~{}lewis.

\bibitem{Tentyukov:2006ys}
  M.~Tentyukov and J.~A.~M.~Vermaseren,
  Comput.\ Phys.\ Commun.\  {\bf 176} (2007) 385
  [arXiv:cs.sc/0604052].

\bibitem{Chetyrkin:1981qh}
  K.~G.~Chetyrkin and F.~V.~Tkachov,
  Nucl.\ Phys.\  B {\bf 192} (1981) 159.

\bibitem{Bekavac:2009gz}
  S.~Bekavac, A.~G.~Grozin, D.~Seidel and V.~A.~Smirnov,
  Nucl. Phys. B {\bf 819} (2009) 183
  [arXiv:0903.4760 [hep-ph]].

\bibitem{vanRitbergen:1998pn}
  T.~van Ritbergen, A.~N.~Schellekens and J.~A.~M.~Vermaseren,
  Int.\ J.\ Mod.\ Phys.\  A {\bf 14} (1999) 41
  [arXiv:hep-ph/9802376].
\end{thebibliography}
\end{document}